\documentstyle[preprint,aps,tighten]{revtex}

\begin{document}

\baselineskip=14.2pt
\pagestyle{plain}

\renewcommand{\thefootnote}{\fnsymbol{footnote}}

\newcommand{\nc}{\newcommand}
\nc{\grad}{\nabla}
\nc{\tr}{\mathop{\rm tr}}
\nc{\half}{{1\over 2}}
\nc{\third}{{1\over 3}}
\nc{\be}{\begin{equation}}
\nc{\ee}{\end{equation}}
\nc{\bea}{\begin{eqnarray}}
\nc{\eea}{\end{eqnarray}}
\nc{\al}{\alpha}
\nc{\ga}{\gamma}
\nc{\de}{\delta}
\nc{\ep}{\epsilon}
\nc{\ze}{\zeta}
\nc{\et}{\eta}
\renewcommand{\th}{\theta}
\nc{\Th}{\Theta}
\nc{\ka}{\kappa}
\nc{\la}{\lambda}
\nc{\rh}{\rho}
\nc{\si}{\sigma}
\nc{\ta}{\tau}
\nc{\up}{\upsilon}
\nc{\ph}{\phi}
\nc{\ch}{\chi}
\nc{\ps}{\psi}
\nc{\om}{\omega}
\nc{\Ga}{\Gamma}
\nc{\De}{\Delta}
\nc{\La}{\Lambda}
\nc{\Si}{\Sigma}
\nc{\Up}{\Upsilon}
\nc{\Ph}{\Phi}
\nc{\Ps}{\Psi}
\nc{\Om}{\Omega}
\nc{\ptl}{\partial}
\nc{\del}{\nabla}
\nc{\ov}{\overline}
\nc{\ads}{AdS$_3\,$}
\nc{\bi}{\bibitem}

\newcommand{\ww}{\wedge} 
\newcommand{\pp}{\partial }

\draft
\preprint{
\vbox{\hbox{DAMTP-2002-20}
      \hbox{hep-th/0203021}}}
\title{Holography and the Polyakov action}

\author{M. Ba\~nados$^1$\footnote{email: {\tt mbanados@fis.puc.cl}}, 
O. Chand\'{\i}a$^1$\footnote{email: {\tt
ochandia@maxwell.fis.puc.cl}}, and 
A. Ritz$^2$\footnote{email: {\tt a.ritz@damtp.cam.ac.uk}}}

\address{$^{(1)}$Departamento de F\'{\i}sica, P. Universidad Cat\'olica 
de Chile, \\ Casilla 306, Santiago 22, Chile \\
$^{(2)}$Department of Applied Mathematics and Theoretical
Physics,  \\ Centre for Mathematical Sciences,
University of Cambridge, \\ Wilberforce Rd., Cambridge CB3 0WA, UK}

\maketitle
\thispagestyle{empty}
\setcounter{page}{0}

\begin{abstract}

In two dimensional conformal field theory the generating functional
for correlators of the stress-energy tensor is given by the non-local
Polyakov action associated with the background geometry. We
study this functional holographically by calculating the regularized
on-shell action of asymptotically AdS gravity in three dimensions, associated
with a specified (but arbitrary) boundary metric. This procedure
is simplified by making use of the Chern-Simons formulation, and 
a corresponding first-order expansion of the bulk dreibein, rather
than the metric expansion of Fefferman and Graham. The dependence
of the resulting functional on local moduli of the boundary metric
agrees precisely with the Polyakov action, in accord with the AdS/CFT
correspondence. We also verify the consistency of this result with
regard to the nontrivial transformation properties of bulk solutions
under Brown-Henneaux diffeomorphisms.

\pacs{}

\end{abstract}

\section{Introduction}

The interplay between two-dimensional conformal field theories
and classical three-dimensional gravity with a negative 
cosmological constant can be traced back to the identification, by Brown
and Henneaux \cite{BH}, of an infinite-dimensional symmetry acting
on the space of gravitational solutions asymptotic to 
anti-de Sitter space (\ads). On the two-dimensional
conformal boundary at infinity this symmetry reduces to two copies 
of the Virasoro algebra with central charge,
\be
 c  =  \frac{3l}{2G_3}, \label{cBH}
\ee
where $l$ is the AdS$_3$ scale (set to unity from hereon), 
and $G_3$ is the three-dimensional Newton constant. This structure,
now embedded within the general AdS/CFT correspondence
\cite{Maldacena,Gubser-,Witten,adsrev}, has come under considerable 
recent scrutiny. In particular, Strominger's observation 
\cite{Strominger97} that a unitary CFT on the boundary, with central
charge (\ref{cBH}), would have a density of states sufficiently large 
to account for the entropy of 3D BTZ black holes \cite{BTZ,BHTZ}, 
has stimulated further work on particular realizations of this system 
in string theory, such as configurations of fundamental strings 
and NS5-branes wrapped on either $T^4$ or $K3$, in the 
hope of understanding the dual CFT \cite{sv} and consequently the 
microscopic origin of the black hole entropy.
However, string theory on noncompact target spaces such as \ads
is still rather mysterious \cite{pmp} (see e.g. \cite{brfw,su11} for
earlier work), and recent studies \cite{mo} 
(see also \cite{GKS,BORT,bvw,KS,SW,br,bars,GK}) are only now leading
in particular cases to a consistent picture of the perturbative spectrum.

In this context, it is interesting to explore how information
about the dual CFT is encoded at the purely gravitational
level, namely in the space of solutions to asymptotically AdS 3D
gravity. All such solutions are locally anti-de Sitter, and starting
from pure \ads, the classical phase space may be constructed in terms
of orbits of the Brown-Henneaux mapping. Recall, for example, 
that a solution of the form,
\be
 ds^2_{3D} = dr^2 + e^{2r}dz d\bar z + \frac{12\pi}{c} T(z)dz^2
+\cdots, \label{chiralmet}
\ee
asymptotic to \ads for large $r$, where $r$ is the radial coordinate, 
admits infinitesimal Brown-Henneaux diffeomorphisms \cite{BH} 
(with parameter $\ep(z)$) as an asymptotic symmetry under which 
the metric remains form-invariant, up to the shift
\be
 \de T = \ep\ptl T + 2T\ptl \ep -\frac{c}{24\pi}\ptl^3\ep, \label{inf}
\ee
where $\ptl=\ptl/\ptl z$. These mappings reduce on the boundary
to infinitesimal local conformal transformations, and we see from (\ref{inf})
that in accord with the AdS/CFT correspondence we can identify the 
subleading components $T(z)$ (and $\bar T(\bar z)$) of the metric 
(\ref{chiralmet}) with the expectation value of chiral (and
anti-chiral) components the boundary stress-energy 
tensor \cite{martinec,Navarro-N,banados99,adsrev,SS,bers,eff}.
Extending consideration to {\it finite} Brown-Henneaux mappings,
relates solutions with different topology, given suitable
coordinate identifications. However, in practice it is convenient to
consider orbits of fixed topology generated by (\ref{inf}) which
are characterized by a given background  topology (say pure \ads),
perturbed by Brown-Henneaux `gravitational waves' (described by $T$) in 
(\ref{chiralmet}).

This picture of the classical phase space generalizes to
solutions with a specified conformal structure $[g]$ at infinity,
which then depend on a general representative metric $g_{ij}$ in this
conformal equivalence class (in (\ref{chiralmet}) $g_{ij}$ is simply
the flat metric). These bulk geometries may be obtained via the
construction of Fefferman and Graham \cite{FG} (see also
\cite{Henningson-S}), and take the
asymptotic form \cite{eff,bers,RS1,K2},
\be
 ds^2_{3D} = dr^2 + e^{2r}g_{ij}dx^idx^j + \frac{1}{2}\left( {\cal R}
g_{ij} + \frac{24\pi}{c} \left< T_{ij}\right>\right)dx^idx^j 
 + \cdots, \label{FGgen}
\ee
where $\langle T_{ij}\rangle$, which is traceless, is unconstrained by the
bulk Einstein equations \cite{FG}. This is consistent with its
anomalous transformation under Brown-Henneaux diffeomorphisms
(as in (\ref{inf})), and 
$\langle T_{ij}\rangle$ can again be identified with
the expectation value of the boundary stress-energy tensor
\cite{eff,bers,RS1,K2}. Thus, in general, 3D bulk Einstein 
solutions can be reconstructed given two pieces of 
boundary data, $\{g,\langle T\rangle\}$. In addition, one has 
holonomy data which describe the global properties\footnote{Such
global data needs to be specified in prescribing $\langle T\rangle$
since the stress-energy tensor undergoes Casimir-type shifts 
on changing the topology.} of the 3D geometry.
 
The possibility of turning on sources $g_{ij}$ for the 
stress-energy tensor $T_{ij}$ in the boundary CFT
allows consideration of how the bulk dynamics reproduces the current
sector of the CFT associated with correlators of the stress-energy
tensor. This is the topic we will now focus on, and we aim in 
the present paper to tackle the local part of this problem,
which may also be inverted as the question of how the 
bulk theory encodes (holographically) the local geometric moduli
of the boundary CFT. In this context the basic quantity  
in the CFT is the generating functional $W[g]$ for correlators 
of the stress-energy tensor,
\be
e^{iW[g_{ij}]} \equiv 
 \left< \exp\left[i\int_{\Si} g^{ij}T_{ij}\right] \right>_{CFT}
\ee 
where $\{\Si,g_{ij}\}$ is the conformal boundary of the asymptotically AdS
bulk geometry. This boundary coupling leads to a Weyl anomaly given by
\be
 \left< T^i_i \right>_{CFT} = -\frac{c}{24\pi} {\cal R}, \label{weyl}
\ee   
where ${\cal R}$ is the scalar curvature associated with the metric 
$g_{ij}$, and $c$ is the corresponding central charge. Using this,
the anomalous Ward identity for Weyl transformations can be integrated
\cite{Polyakov} leading to the Polyakov action as the expression
for the generating functional $W[g]$ in covariant form,
\be
 W_P[g] = {c \over 96\pi} \int \int\left[ {\cal R}\, {1 \over \nabla^2} \,
{\cal R} + \la^2\right],
\label{Polyakov}
\ee
where $\la^2$ is a cosmological constant.

From the bulk point of view, the AdS/CFT correspondence implies 
a relation between the generating functional $W[g]$ and the
string partition function evaluated with conformal boundary data
$g_{ij}$. In the classical gravitational regime, this relation
takes the form 
\cite{Maldacena,Gubser-,Witten,adsrev,eff}\footnote{In general, it may be 
necessary to specify topological data, $\ga$, and to sum over
inequivalent bulk topologies having the same boundary
\cite{witten2}, although we will not need to consider the latter cases
here.}
\be
 W[g] \sim I_{\rm reg}[G] = \lim_{\ep \rightarrow 0} 
  (I_{\rm EH}[G_{\ep}] - I_{\rm ct}[g_{\ep}]), \label{screln}
\ee
where $I_{\rm EH}[G_{\ep}]$ is the bulk Einstein-Hilbert action (plus
the appropriate boundary terms) evaluated on a solution $G_{ij}[g]$ 
of the form (\ref{FGgen}) with boundary conformal structure $[g]$, 
and an infrared regulator $r<1/\ep$ is used to subtract the 
bulk divergences with covariant counterterms $I_{\rm ct}$ 
\cite{Henningson-S,Balasubramanian-K}
(for further work on holographic renormalization see \cite{SS,eff}). 

The duality in (\ref{screln}) has already been
sucessfully tested in the computation of Weyl anomalies 
of the boundary CFT via regularization of the bulk action. 
The calculation outlined in  \cite{Witten}, and 
carried out explicitly by Henningson and Skenderis 
\cite{Henningson-S} makes use of diffeomorphism invariance 
in the bulk. Specifically, by looking at 
the logarithmically divergent terms in the Einstein-Hilbert action 
evaluated on a general asymptotically AdS solution \cite{FG}, the 
variation of the action under Weyl transformations can be computed 
and the expected expressions for Weyl anomalies in various 
dimensions were obtained \cite{Henningson-S}. In particular, for
$d=2$, the result (\ref{weyl}) was reproduced with a central charge
given by (\ref{cBH}), consistent with the expected Weyl 
anomaly for a CFT realizing the Brown-Henneaux \cite{BH} asymptotic 
conformal symmetry.

The correspondence (\ref{screln}) is, however, considerably stronger
than just a relation between the Weyl anomalies as it implies a
direct equivalence between the generating functionals. In this
paper, we will focus on the induced boundary effective action 
and its relation to the generating functional\footnote{See
\cite{SS,eff,K2} for other work on the induced
effective action via AdS/CFT in various dimensions.} $W[g]$ given
in (\ref{Polyakov}). An advantage of working within the \ads/CFT$_2$ 
framework is that the dependence of $W_P[g]$ (\ref{Polyakov}) on 
the underlying moduli of the metric on $\Si$ is well-understood, 
and this can be contrasted with the corresponding dependence of 
the bulk action $I_{\rm reg}[g]$. Working within the Chern-Simons formalism,
we will show that this moduli dependence is precisely that of the 
Polyakov action (\ref{Polyakov}), a relation that has also been 
argued to hold by Skenderis and Solodukhin \cite{SS} using a different
approach.

The appearance of the Polyakov
functional through the relation (\ref{screln}) may be anticipated
by observing that, from the point of view of the boundary CFT, 
the right hand side of (\ref{screln}) represents a particular 
covariant 3D `localization' of the generically non-local generating 
functional $W[g]$. To interpret this it is helpful to recall 
that the standard means of localizing the Polyakov action 
involves the introduction of a Liouville field $\ph$. If the background metric 
on $\Si$ is $g_{ij}$, then we consider a new background
$e^{\ph}g_{ij}$ with constant curvature. The Liouville
action (dropping the potential) for $\ph$ is then given by,
\be
 W_L[g,\phi] = -\frac{c}{48\pi} \int_{\Si} d^2 z \sqrt{-|g|}\left[
 \frac{1}{2}g^{ij}\del_i\phi \del_j \phi + \phi {\cal R}[g]\right],
   \label{WL}
\ee
which, on integrating out $\phi$, reduces to the non-local Polyakov 
action (\ref{Polyakov}) for $g_{ij}$. This procedure can be 
realized within the AdS/CFT correspondence if we consider the dynamics 
on a regulated boundary $\Si_{r_0}$ at fixed radial coordinate $r=r_0$
in the asymptotic regime. We can regard the radial dependence 
as described by a field $r_0=r_0(\Si_{r_0})$. It follows
from the form of the asymptotically AdS bulk metric (\ref{FGgen}) that 
under local conformal transformations on $\Si_{r_0}$, 
$r_0$ transforms as a Liouville field. This identification of the bulk 
radial coordinate with a Liouville field \cite{polyakov97} is simply the
standard UV/IR correspondence \cite{PP} with radial shifts 
translated to scale transformations on the boundary. The corresponding
behaviour of regulating surfaces was discussed in some detail in 
\cite{K1} (see also \cite{SW}). This picture provides qualitative
evidence for the appearance of the Polyakov action as the
boundary generating functional. The main aim of this paper will be
to verify this relation in detail.

Following the same theme, there is in the present context
another aspect of the bulk `localization' which will be of 
interest. Recall that the Polyakov action
evaluated in a light-cone gauge background was originally constructed
\cite{Polyakov} as a gravitational WZW model, and thus has a natural
local representation in 3D. This is the appropriate background
geometry on $\Si$ in which to make explicit the dependence on `complex'
structure moduli and we will find that there is a nice mapping
between these chiral and anti-chiral moduli and the degrees of
freedom entering via the two Chern-Simons fields which arise in the 
first-order formulation of the bulk dynamics. This in part motivates 
our approach to the calculation of the bulk action in (\ref{screln})
which makes use of the Chern-Simons formalism. The relevance of this
formalism for realizing the holomorphic factorization associated 
with the generating functional of a CFT was also discussed recently
in \cite{K2}.

An important issue which arises in interpreting (\ref{screln}) 
is that fixing the boundary conformal structure $[g_{ij}]$ does
not in general specify a unique bulk continuation \cite{FG}, due to the 
anomalous transformation properties of $\langle T \rangle$, 
as discussed above. This concerns the first variation of (\ref{screln}),
\be
 \langle T \rangle_{\rm CFT} \sim \frac{1}{\sqrt{-|g|}}
\frac{\de I_{\rm reg}[g]}{\de g}, \label{Treln}
\ee
and we will show that bulk Brown-Henneaux transformations
which shift $\langle T \rangle$ correctly maintain the
correspondence  in (\ref{Treln}). Specifically, the
ambiguities associated with bulk Brown-Henneaux
diffeomorphisms enter, as expected, via local conformal transformations
on the boundary. i.e. the moduli only determine the conformal class
of the boundary metric and Brown-Henneaux diffeomorphisms amount to
free-field shifts in the corresponding conformal mode.

In concluding this section, it is worth noting that when
using a conformal gauge for the boundary metric, the Polyakov
generating functional we obtain reduces to that of Liouville theory, where 
the presence or otherwise of the potential term depends on 
the renormalization condition for the two-dimensional cosmological
constant.  However, it is important to realize that the 
Liouville theory found here should not be identified with that 
obtained by Coussaert, Henneaux and van Driel (CHvD) \cite{CHvD} 
describing the asymptotic dynamics of gravity in \ads.  
More precisely, in \cite{CHvD} the Liouville dynamics 
describes gravitational perturbations propagating on a 
fixed 2D background metric $g_{ij}$ (see \cite{HMS,bers,RS1} for 
a generalization of \cite{CHvD} to arbitrary backgrounds). 
Due to the lack of local degrees of freedom in 3D gravity,  
these perturbations actually encode  the entire bulk dynamics 
(up to holonomies), for a given set of boundary conditions. 
In \cite{CHvD} only the constraint equations were solved leading to
dynamical fluctuations on the boundary. In contrast, in our analysis, the
entire set of 3D equations are solved, for a given 2D background metric 
$g_{ij}$, thus obtaining the generating functional 
$W[g_{ij}]$  for stress tensor correlators. For a proper 
comparison the CHvD action should be put on-shell, and the on-shell
correspondence between bulk geometries and Liouville solutions has
been discussed in \cite{martinec,K1}.

The paper is organized as follows. In Section~2 we turn to the general
parametrization of bulk metrics, introducing a first-order 
form for the expansion of Fefferman and Graham
\cite{FG}. We then review in Section~3 how 
the generating functional $W_P[g]$ encodes the local moduli of the boundary
metric in explicit form. In the present context we
will restrict attention to simple boundary topologies, and the moduli
are conveniently encoded in a chiral parametrization of the metric,
the analogue of a Beltrami parametrization for Riemann surfaces.
In Section~4 we derive the dependence of $I_{\rm reg}[g]$ 
on the boundary moduli using a covariant Chern-Simons
construction; the result is consistent with (\ref{Polyakov}).
We also discuss the action of Brown-Henneaux diffeomorphisms
on the space of solutions, and the relation to the expectation value
of the boundary stress tensor. Section~6 contains some 
concluding remarks concerning global data.

\section{Chern-Simons formulation and a first-order expansion} 

To implement the AdS/CFT prescription one needs to reconstruct a 
bulk Einstein metric with negative cosmological constant, given 
a representative $g^{(0)}_{ij}$ of the conformal structure at
infinity. In the Chern-Simons formulation, this problem can be
rephrased in first-order form, where the gauge freedom can be used
to generate a solution in a straightforward manner. Before
describing this, we recall some details of the conventional
metric formalism.

Finding a bulk solution given a fixed boundary conformal
structure $[g^{(0)}]$ is a nontrivial problem in general, although a particular
existence theorem was obtained by Graham and Lee \cite{GL} for the
special case of boundary metrics sufficiently close to the standard
one on the sphere. However, it was shown by Fefferman and Graham \cite{FG}
(see also \cite{Henneaux-T}) 
that an asymptotic expansion near infinity can be constructed starting
from an arbitrary boundary metric. This expansion has the special form, 
\be
 ds^2 = dr^2 + e^{2r} g_{ij}(r,x)dx^idx^j,
\ee
with
\be
 g_{ij}(r,x) = g^{(0)}_{ij} + e^{-2r} g^{(2)}_{ij}+\cdots,  
\ee
where $r$ is a radial coordinate, related to that used in
\cite{FG} by $e^{2r}\sim 1/\rho$. The Einstein equations in general
determine almost all of the coefficients $g^{(n)}_{ij}$ as covariant 
functions of  $g^{(0)}_{ij}$ and its derivatives. The coefficient 
$g^{(0)}_{ij}$, which is defined up to a Weyl rescaling, determines
the boundary conformal structure and is
identified with the boundary metric.

For the 2+1D case of particular interest here, this structure is known
to simplify with the expansion truncating at $O(e^{-4r})$
(for pure Einstein metrics which are all locally AdS). We can 
write \cite{banados99,SS},
\be
 ds^2 = dr^2 + \left(e^{2r}g^{(0)}_{ij} 
 + g^{(2)}_{ij}+ e^{-2r}g^{(2)}_{ik}g^{(2)k}_j\right) dx^i dx^j.
 \label{FGmet}
\ee
Given the boundary metric $g^{(0)}_{ij}$, the
Einstein equations determine the trace of $g^{(2)}_{ij}$
automatically (which is enough to determine the boundary Weyl 
anomaly \cite{Henningson-S}), but do not specify the
trace-free part of $g^{(2)}_{ij}$ \cite{FG,Henningson-S}. 
This `ambiguity' is equivalent to the
choice of a quadratic form\footnote{It has recently been emphasized
that this ambiguity is equivalently understood in Euclidean signature
as a choice of projective structure on the boundary \cite{K2}.} 
on the boundary 
\cite{banados99}, which as noted in the introduction transforms
anomalously under Brown-Henneaux diffeomorphisms. This is consistent 
with the corresponding transformation of the 
boundary stress-tensor under local conformal mappings, and these
quantities are identified via the AdS/CFT correspondence 
\cite{eff,bers,RS1,K2}. We will return to this issue in Section~IV.

A Fefferman-Graham-type expansion can also be formulated in 
first-order form in terms of connections. Recall that a
3D Lorentzian geometry can be written in terms of two flat $SL(2,\Re)$ 
gauge fields $A$ and $\bar A$. i.e. the dreibein and spin-connection 
are given by,
\begin{equation}
e_\mu=(A_\mu-\bar A_\mu)/2  \ \ \ \ \   w_\mu=(A_\mu+\bar A_\mu)/2
\label{ew}
\end{equation}
where $A=A^a J_a$, $\bar A=\bar A^a J_a$. We use the 
$SL(2,\Re)$ basis $\{J_+, J_-,J_3\}$ with $[J_+,J_-] = 2J_3$, 
$[J_3,J_\pm] = \pm J_\pm$ and Tr$(J_+J_-)=1$, Tr$(J_3J_3)=1/2$ .

The Einstein-Hilbert action is then equal to the difference of 
two Chern-Simons actions, supplemented by boundary terms which 
depend on the boundary conditions. Specifically, if we normalize the
Einstein action as
\be 
 I_{EH} = \frac{1}{16\pi G} \int \sqrt{-|g|}(R-2\La) + \mbox{boundary
terms},
\ee
with $\La=-1$ (in units where $l=1$), then
\be
 {4\pi \over k}I_{EH} = I_{CS}[A] - I_{CS}[\bar A] + \mbox{boundary terms},
\ee
where $k=1/4G$ is the level, and $I_{CS}[A]$ is the Chern-Simons functional, 
\begin{equation}
I_{CS}[A] = \int_M AdA + {2 \over 3} A^3,
\label{CS}
\end{equation}
with $M$ the bulk 3D manifold.
We will fix the required boundary terms once we have considered the asymptotic
form of the on-shell connection in comparison with the Fefferman-Graham
expansion in (\ref{FGmet}).

Recall that the space of solutions of three-dimensional 
Chern-Simons theory is the set of
flat SL(2,$\Re$) connections $A = g^{-1}dg$ plus holonomies. 
We shall consider the pure AdS case with no holonomies, although we
will comment on their inclusion in Section~V. Note that  
the breakdown of gauge invariance at the
boundary nonetheless prevents us from setting $g=1$.

The radial dependence of $g$ is itself given by a gauge
transformation, and an appropriate asymptotic radial coordinate can be 
introduced as follows \cite{B,CHvD}, 
\begin{eqnarray}
A = e^{-r J_3} \, \alpha \, e^{r J_3} + dr J_3 \label{alpha} \\
\bar A = e^{r J_3} \, \bar\alpha \, e^{-r J_3} - dr J_3 \label{alphabar} 
\end{eqnarray}        
where $\alpha$ and $\bar\alpha$ are both $SL(2,\Re)-$valued flat
connections defined on the surface at fixed\footnote{In general, to avoid
singularities at $r=0$, and hence the introduction
of other degrees of freedom, one must allow $\alpha$ and $\bar\alpha$
to depend on $r$ near the horizon. We ignore this subtlety here as we
are concerned with the asymptotics near $r\rightarrow \infty$.} $r$.

Expanding $\alpha$ and $\bar \alpha$ in the  basis $J_+, J_-,J_3$, 
we obtain the series expansion for the 3D forms $A$ and $\bar A$, 
\begin{eqnarray}
A  &=& (dr + \alpha^3) J_3 +   e^{-r }\, \alpha^+ J_+ + e^{r }\,
          \alpha^- J_-,
          \label{A2}\\
\bar A  &=&  (-dr +\bar \alpha^3) J_3 +  e^{r }\, \bar \alpha^+ J_+ 
 + e^{-r }\, \bar \alpha^- J_-,
\label{Ab2}  
\end{eqnarray}   
where we have used $e^{-r J_3} J_\pm e^{r J_3} = e^{\mp r} \,
J_\pm$. This series is consistent with the structure
of the Fefferman-Graham  expansion 
(\ref{FGmet}), and the corresponding dreibein 
$e=(A-\bar A)/2$ becomes, 
\begin{eqnarray}
e & = & \left( dr  + {1 \over 2}(\alpha^3 - \bar \alpha^3)\right)J_3 
    + {1 \over 2} e^r ( \alpha^- J_- -\bar \alpha^+J_+)  
    + {1 \over 2} e^{-r} ( \alpha^+ J_+ -\bar \alpha^- J_-).
\label{e}
\end{eqnarray}

In analogy with the metric treatment, we define the
conformally induced boundary 2D zweibein as the leading term in the 
$r\rightarrow \infty $ limit,  
\begin{equation}
e^{(0)} = \frac{1}{2}\left( \alpha^- J_- -\bar \alpha^+J_+ \right), 
\label{bzweibein}
\end{equation}
where $\Si=\ptl M$ is identified as the conformal boundary of the
Poincar\'e patch, and not the global boundary of \ads.
Note that this procedure yields a completely general zweibein on
the surface $\Si$,  which is determined by two independent 1-forms
from $\alpha$ and $\bar\alpha$. Note, however, 
that of the four components entering the zweibein, only three are 
(locally) independent as one may be decoupled in 
the line element by 2D Lorentz transformations. 

At this point, it is clear that since we wish to fix the conformal
structure at the boundary, it is necessary to impose suitable
(conformal) Dirichlet conditions on $\al^-$ and $\bar\al^+$. Actually,
since we want to fix the conformal structure, but not a particular
representative, we write the 
corresponding condition in terms of $A$ and $\bar A$,  
\begin{equation}
\delta ( A-\bar A)^+ = 0, \ \ \ \ \   \delta ( A-\bar A)^- = 0
\label{bc}
\end{equation}
The remaining boundary condition on $A^{3}$ and $\bar A^3$ follows 
by comparing the metric determined by (\ref{e}) with the block-diagonal
metric of the Fefferman-Graham expansion (\ref{FGmet}). We see that
the off-diagonal term $dr (\alpha^3_i-\bar \alpha^3_i)dx^i$ arising 
from (\ref{e}) is canceled in the asymptotic regime on imposing the 
`Neumann' boundary condition, 
\begin{equation}
\alpha^3 - \bar \alpha^3  = 0.
\label{Neumann}
\end{equation}
We will see later that this condition is also sufficient to ensure
that the boundary metric is torsion-free. These boundary conditions
lead to a 3D geometry parametrised by the metric
(following from (\ref{e})),
\be
 ds^2 = dr^2 - e^{2r}\al^-\bar\al^+ 
+ (\al^-\al^++\bar\al^+\bar\al^-)-e^{-2r}\al^+\bar\al^-,
         \label{fullmet}
\ee
which we see is consistent with the truncated expansion 
in (\ref{FGmet}).

The appropriate action for imposing Dirichlet conditions on 
$(A-\bar A)^\pm$ and a Neumann condition on $(A- \bar A)^3$ 
is\footnote{A comparison with the dreibein formulation may be useful here. 
The 3D Einstein-Hilbert action written in frame variables 
is $\int R_a \wedge e^a$ 
where $R^a = dw^a +(1/2) \epsilon^a_{\ bc}w^b\wedge w^c$.  
Varying this action one picks up the boundary term 
$\int \delta w_a \wedge e^a$ and thus either a Dirichlet condition on 
$w^a$ or a Neumann condition on $e^a$ is required. Conversely, one can
write an action appropriate for Dirichlet conditions on $e^a$ or 
Neumann conditions on $w^a$ by adding the boundary term 
$-\int w_a \wedge e^a$.   In the Chern-Simons formulation we consider 
the action $I_\pm=I_{CS}[A] - I_{CS}[\bar A] \pm \int A \wedge \bar A$
whose variation yields the boundary term 
$\int (A \pm \bar A)\wedge\delta (A \mp \bar A) $.  The sign has to be
chosen according to whether we want to fix the connection, $A+\bar A$,
or the dreibein, $A-\bar A$, and to whether these conditions are
Dirichlet or Neumann.  In our situation, we have a mixed case with 
Dirichlet conditions on $(A-\bar A)^\pm$  and Neumann conditions 
on $(A-\bar A)^3$. This leads directly to the action (\ref{3dI}). }
\begin{eqnarray}
\frac{4\pi}{k}I & = & I[A] - I[\bar A] + 
 \int_{\partial M} ( A^+ \wedge \bar A^- + A^- \wedge 
\bar A^+ - {1 \over 2} A^3 \wedge \bar A^3).
\label{3dI}
\end{eqnarray}
The negative sign in front of the term $A^3\wedge \bar A^3$ is
required to impose the Neumann condition on $A^3 - \bar A^3$.
In other words, the variation of the action (\ref{3dI}) will have a
term $ \int_{\pp M} (A^3-\bar A^3) \delta(A^3+\bar A^3)$.  We demand
the action to be stationary with respect to arbitrary variations of 
$A^3+\bar A^3$ at the boundary. The condition $\al^3-\bar \al^3=0$ then 
follows as an equation of motion in the boundary theory.  

It is important to note that this treatment has been 
manifestly covariant on the 2D surface $\Si$. In particular, 
we have not needed any coordinate choice to fix the boundary terms.

\section{Boundary Moduli and the Polyakov Action}

Before turning to the analysis of the bulk action (\ref{3dI}),
we consider the expected form of the boundary generating functional
(i.e. the Polyakov action) in suitable test geometries on $\Si$. 
The boundary metric (\ref{bzweibein})
determined in the previous section is sufficiently general to 
allow an arbitrary dependence on local moduli. 

To make this dependence manifest, we first recall that in 2D 
any metric is locally conformally flat, and thus can be represented
in suitable coordinates as
\be
 ds^2 = e^{\varphi'} dx d\bar x, \label{cflat}
\ee
where $\varphi'(x,\bar x)$ is the conformal mode. We are considering 
boundary geometries with Lorentzian signature, and so $\{x,\bar x\}$
are independent real light-cone coordinates, although the
transition to Euclidean signature (for low genus) will be clear
from the notation. 

The parametrization
(\ref{cflat}), while simple, hides the dependence on the `complex'
structure of the surface $\Si$. To make this manifest, we consider
a quasi-conformal mapping $\{x,\bar x\} \rightarrow \{z(x,\bar x),\bar
z(x,\bar x)\}$, where $x$ and $\bar x$ satisfy the equations,
\be
 \bar \ptl_{\bar z} x = \mu \ptl_z x, 
 \;\;\;\;\;\;\;\;\;\;\;
 \ptl_z \bar x = \bar \mu \bar \ptl_{\bar z} \bar x. \label{beltrami}
\ee
where $\mu$ and $\bar \mu$ are independent, real and
bounded ($|\mu|$ and $|\bar\mu|<1$) functions, and are 
the (Lorentzian) analogue of Beltrami parameters. 
The metric (\ref{cflat}) then takes the form
\be
 ds^2 = e^{\varphi} (dz + \mu d\bar z) (d\bar z + \bar \mu dz). \label{exp}
\ee
We will restrict our attention 
to boundaries of cylindrical topology, and thus 
this parametrization is sufficient to describe an arbitrary metric on
$\Si$ given a fixed coordinate system $\{z,\bar z\}$. The metric
(\ref{exp}) also makes explicit the dependence on a representative of the 
conformal class ($\varphi$), and on the deformations of the 
`complex' structure $(\mu,\bar\mu)$ specifying the conformal class. 
These three moduli map to the
three (independent) components of the boundary zweibein
(\ref{bzweibein}).

We will find it useful to consider particular examples of the
general boundary geometry (\ref{exp}). Recalling that conformal symmetry in 2D 
is generated by chiral and anti-chiral stress tensors $T(z)$ and
$\bar T(\bar z)$, it will be sufficient to consider the dependence on the 
conformal mode $\varphi$ and one chiral parameter $\mu$. These
components of the metric are sources for the stress-energy tensors
$T^{\varphi}(z,\bar z)$ and $T(z)$ respectively. Note that the distinction 
here is tied up with the question of whether 
(\ref{beltrami}) has a global solution, or in other words whether
or not $ds^2$ and the metric 
$d\tilde{s}^2 = \exp(\varphi)dz d\bar z$ are in the same conformal
class (see e.g. \cite{DHP}).

With this information in hand, we can determine the form of the 
Polyakov generating functional $W_P[\varphi,\mu,\bar\mu]$ 
which exhibits the explicit dependence on the moduli. The general 
decomposition of $W_P$ for the metric (\ref{exp}) was obtained by 
Verlinde \cite{verlinde}, but for simplicity we will restrict our attention
to boundary geometries depending either on $\varphi$ or $\mu$.

\subsection{Conformal Gauge}

Consider first the conformal gauge where, with a signature convention 
chosen for later convenience, the metric takes the form,
\be
 d\tilde{s}^2 = - \exp(\varphi)dz d\bar z. \label{cflatm}
\ee
The corresponding zweibein (\ref{bzweibein}) is given by,
\be
 e^{(0)} =  \frac{1}{2}\left(\alpha^- J_- -\bar \alpha^+J_+\right) = 
 \frac{1}{2}\left(e^{\ph}dzJ_- - e^{\bar
\ph}d\bar z J_+\right), \label{czwei}
\ee
where $\varphi=\ph+\bar\ph$, and we see that in general the conformal
mode will receive contributions from both gauge fields $A$ and $\bar
A$ entering the bulk action.
In this background the Polyakov action reduces to that of Liouville theory,
\be
 W_P[\varphi] = -\frac{c}{96\pi}\int_{\Si} d^2 z
\left[\ptl\varphi\bar\ptl\varphi + \la^2e^{\varphi}\right]. \label{conformal}
\ee
where $\ptl=\ptl/\ptl z$, and $\bar\ptl=\ptl/\ptl\bar z$. There is
clearly no curvature coupling as the reference background in this case
(\ref{cflatm}) is flat.

\subsection{Light-cone gauge}

If instead we consider Polyakov's light-cone gauge, with
\be
 d\tilde{s}^2 = -dz d\bar z - \mu d\bar z^2, \label{lcmet}
\ee
the corresponding zweibein (\ref{bzweibein}) is given by,
\be
 e^{(0)} =  \frac{1}{2}\left(\alpha^- J_- -\bar \alpha^+J_+\right) = 
 \frac{1}{2}(dz+\mu d\bar z)J_- - \frac{1}{2}d\bar z J_+,
    \label{lczwei} 
\ee
which makes the chiral structure quite manifest. We see that the
nontrivial dependence on the `Beltrami' parameter $\mu$ enters only
via $A$ and not $\bar A$. In contrast, it is clear that in an 
anti-chiral gauge, the dependence on $\bar \mu$ would enter via $\bar
A$. Recalling that $\mu$ and $\bar\mu$ act as sources for the chiral
and anti-chiral stress tensors $T$ and $\bar T$, which generate the
two copies of the Virasoro algebra comprising the conformal
group, this correspondence provides a simple map from 
`holomorphic' factorization on the boundary to the obvious factorization
of the two Chern-Simons systems in the bulk \cite{K2}.

This conclusion is perhaps not as obvious as it might seem due to the
collapse of the bulk dynamics to the boundary, where the boundary terms
couple $A$ and $\bar A$. What we observe from (\ref{czwei}) and
(\ref{lczwei}) is that as expected the conformal mode reflects 
a violation of this factorization, while the light-cone gauge is
special in that it is preserved. We might anticipate that the
simple factorization observed in the light-cone gauge is related to
Polyakov's observation (see also \cite{verlinde}) 
of an SL(2,$\Re$) structure in 2D gravity.
However, the fact that (\ref{lczwei}) still involves both gauge fields
$A$ and $\bar A$ makes the relation 
obscure\footnote{Note that the geometric data for $\Si$ can be
combined to form a {\it single} flat SL(2,$\Re$) connection
($A = -i\om J_3 + e^+J_+ + e^- J_-$), where the constraint
$dA + A \wedge A = 0$ arises directly in Hamiltonian
Chern-Simons theory.}, and we will not explore
this issue further here.

Returning to the generating functional, when evaluated in the 
background (\ref{lczwei}), one obtains the light-cone gauge 
Polyakov action
\cite{Polyakov} (see also \cite{KPZ})
\be
 W_P[\mu] = -\frac{c}{48\pi} \int_{\Si} d^2 z \frac{\ptl^2 x}{\ptl x}
\ptl\mu,
 \label{chiral}
\ee
where we have ignored the (constant) potential term and
$x=x(z,\bar z)$ satisfies the equation (\ref{beltrami}).

\section{The holographic generating functional}

We now return to the bulk action (\ref{3dI}), and determine its
dependence on the boundary moduli for comparison with
(\ref{conformal}) and (\ref{chiral}).
The bulk equations of motion are simply $d\alpha + \alpha\wedge
\alpha=0$  and similarly for $\bar \alpha$, and it is useful to write 
these equations explicitly in the SL(2,$\Re$) basis, 
\begin{eqnarray}
d\alpha^3 + 2 \alpha^+\ww \alpha^- = 0,  &&  \;\;\;\;\;\;\;\;\; 
 d\bar\alpha^3 + 2 \bar\alpha^+\ww \bar\alpha^- = 0          \label{flat1} \\
d\alpha^- - \alpha^3 \ww \alpha^- = 0,   &&  \;\;\;\;\;\;\;\;\; 
 d\bar\alpha^+ + \bar\alpha^3 \ww \bar\alpha ^+ =0           \label{flat2} \\
d\alpha^+ + \alpha^3 \ww \alpha ^+ =0,    &&  \;\;\;\;\;\;\;\;\;
 d\bar\alpha^- - \bar\alpha^3 \ww\bar \alpha^- = 0            \label{flat3} 
\end{eqnarray}
The components of the zweibein $\alpha^-$ and $\bar\alpha^+$ appear in
(\ref{flat2}).  In view of (\ref{Neumann}), the pull-back of 
Eq.~(\ref{flat2}) to the boundary can be regarded as the torsion-free
condition for the spin connection coefficient $w \equiv \alpha^3$, with 
associated 2-form curvature  ${\cal R}=dw$, and this observation will 
be useful in what follows.

The on-shell value of the Einstein-Hilbert action can be obtained
straightforwardly by substituting (\ref{A2}) and (\ref{Ab2}) in
(\ref{3dI}). We obtain (suppressing the $\wedge$-product),  
\begin{eqnarray}
{4\pi \over k}I &=& -{1 \over 3} \int_{M} {\rm tr}\left[ (\alpha)^3 
- (\bar\alpha)^3 + 3((\alpha)^2 + (\bar \alpha)^2) J_3dr\right]\nonumber\\ 
 && \;\;\;\;\;\; + \int_{\partial M} e^{2r} \alpha^- \bar\alpha^+ 
 +  e^{-2r} \alpha^+ \bar\alpha^-  
 - {1 \over 2} \alpha^{3} \bar\alpha^{3}.  
\label{on-shell}
\end{eqnarray}
As expected, this expression contains finite terms along with
a quadratic (in $e^{r}$) and logarithmic divergence. 
Note that $\alpha$ and $\bar\alpha$ are still restricted by the
flatness conditions (\ref{flat1},\ref{flat2},\ref{flat3}).

\subsection{Weyl anomaly}

From the on-shell action (\ref{on-shell}), it is straightforward 
to recover the expression for the boundary Weyl anomaly which,
following the discussion of \cite{Witten,Henningson-S}, is given by the
coefficient of the logarithmically divergent term, as all the other
terms are Weyl invariant. The relevant term
is $ \int \mbox{Tr} (\alpha^2+ \bar \alpha^2) \wedge J_3 dr = (1/2)
\int dr \int_{\pp M}  (\alpha^+ \wedge \alpha^- + \bar \alpha^+\wedge
\bar\alpha^- )$.  Using the equations of motion (\ref{flat1}) and the
Neumann boundary condition (\ref{Neumann}) the coefficient reduces to 
$\int_{\partial M} dw = \int dr \int_{\partial M} {\cal R}$, yielding the 
Weyl anomaly,
\be
 \left< T^i_i\right> = -\frac{c}{24\pi} {\cal R}^{(0)}, \label{weylA}
\ee
in agreement with the calculation in \cite{Henningson-S}, 
where ${\cal R}^{(0)}$ is the boundary Ricci scalar,
and we have used $c=6k$.

\subsection{Boundary generating functional}

The other divergent term (which is Weyl invariant) is 
$ e^{2r} \int \alpha^- \wedge \bar\alpha^+$.  This contribution is 
precisely the cosmological constant term in two dimensions because
 $ \alpha^- \wedge \bar\alpha^+ $ is equal to the determinant of the 
zweibein (\ref{bzweibein}).  The appearance of this divergence is 
consistent with the analysis of \cite{Balasubramanian-K}, and it can
be associated with a divergent bare cosmological constant in the CFT. 
Therefore, we shall renormalize this term, canceling the divergence 
but retaining a finite cosmological constant $\la^2$. As in 
2D quantum gravity, we find that the Liouville potential 
comes from this term.

The finite part of the action is therefore, 
\begin{equation}
{4\pi \over k}I_{reg} = 
 -\int_M (\alpha^3\al^+\al^- - \bar\alpha^3 \bar\al^+\bar\al^-) 
 - {1 \over 2} \int_{\pp M} (\alpha^{3} \bar\alpha^{3} + 
 2\la^2 \alpha^-\bar\alpha^+).  
\label{Ifinite}
\end{equation}
and, since $\alpha=g^{-1}dg$ and $\bar\alpha = \bar g^{-1}d\bar g$, it
only depends on the boundary values of the fields. Note that in 
general the group elements $g$ and $\bar g$ are not single-valued.

Localizing the action (\ref{Ifinite}) in general
requires a parametrization
of the holonomies of the gauge fields, thus specifying the topology in
terms of the conjugacy classes of SL(2,$\Re$). We will return to this
issue in Section~V, but as mentioned earlier it will not be 
necessary to explicitly specify a particular class as we are interested 
in the dependence of $I_{reg}$ on local quantities. 
It will be sufficient here to choose a suitable 
local patch of \ads, which is consistent with a choice of boundary
coordinates.

\subsubsection{Boundary conformal gauge}

The simplest choice is one in which
the metric may be written in conformally flat form. The 
appropriate boundary zweibein (\ref{czwei}) is given by
\be
 e^{(0)} = \frac{1}{2}\left( e^{\ph}dx J_- - e^{\bar \ph}d\bar x
J_+\right), \label{zweibein}
\ee
where the conformal mode of the metric is then $\varphi=\ph+\bar\ph$.
Recalling the discussion of Section~III, this coordinate choice
hides the moduli in the conformal mode $\varphi$, as in (\ref{cflat}).
This coordinate system is convenient for deriving the effective
action, and the dependence on moduli can then be made explicit by
performing a quasi-conformal transformation.

This choice of the boundary metric may be achieved by making use of 
the following Gauss decomposition of the SL(2,$\Re$) group elements,
\begin{equation}
      g = e^{x J_-} e^{\phi J_3} e^{y' J_+}, \ \ \ \ 
 \bar g = e^{\bar x J_+} e^{-\bar \phi J_3} e^{\bar y' J_-},
\label{Gauss}
\end{equation}  
which yields the following expressions for the Chern-Simons currents,
\begin{equation}
\begin{array}{ll}
\alpha^+ = e^{-\phi} (-y^2 dx +dy) \ & \ 
         \bar \alpha^+ = e^{\bar\phi} d\bar x \\
\alpha^- = e^{\phi} dx \ & \ 
        \bar\alpha^- =e^{-\bar\phi} (-\bar y^2 d\bar x +d\bar y) 
\label{currents} \\
\alpha^3 = -2y dx + d\phi  \ & \ 
    \bar \alpha^3 = 2\bar y d\bar x - d\bar\phi .  
\end{array}
\end{equation}
Here we have made the convenient replacements $y' = e^{-\phi} y$ and 
$\bar y' = e^{-\bar\phi} \bar y$.

The action written in terms of these fields now takes a
local form on $\Si$ given by,
\begin{equation}
 \frac{4\pi}{k} I_{reg}[\varphi,x,\bar x;y,\bar y] 
 = \frac{1}{2}\int 2\varphi (dx \wedge dy - d\bar x \wedge d\bar y) 
 + (4 y\bar y  + \la^2 e^\varphi) dx \wedge d\bar x
\label{Ireg}
\end{equation} 
with $\varphi = \phi+\bar\phi$ the conformal mode in the metric. 
Note that the mode $\phi-\bar\phi$ has decoupled, which is a 
consequence of the Neumann boundary condition (\ref{Neumann}) on the 
spin connection, and our neglect of holonomies (see e.g. \cite{Navarro-N}).    

The variables $y$ and $\bar y$ are auxiliary fields which can be
eliminated by their own equations of motion.  Using the diffeomorphism
invariance of (\ref{Ireg}) to choose $x$ and $\bar x$ as coordinates, 
the action reads, 
\begin{equation}
I_{\rm reg}[\varphi,y,\bar y] 
= {k \over 8\pi}\int d^2x \left[2\varphi (\bar\pp y+ \pp
 \bar y) + 4 y \bar y + \la^2 e^\varphi\right],
\label{Liouville2}
\end{equation}   
where the derivatives are with respect to $x$ and $\bar x$.
The fields $y$ and $\bar y$ can be integrated out and (ignoring
boundary terms) we obtain the
Liouville action for the conformal mode $\varphi$, 
\begin{equation}
I_{\rm reg}[\varphi] = -{c \over 96\pi} \int d^2 x 
\left[\pp \varphi \bar\pp \varphi + \la^2 e^\varphi\right]. \label{Liou}
\end{equation} 
where we have used $k=1/(4 G )$ and the value of $c$ given 
in (\ref{cBH}). We see that this agrees with the evaluation of the
Polyakov action in the background (\ref{zweibein}), as given in
Section~III. It is also straightforward to check that the 
constraint equations for  $y= \ptl\varphi/2$ and 
$\bar y = \bar\ptl \varphi/2$, 
do imply $\alpha^{(3)} = \bar \alpha^{(3)}$, as required by the
3D variational principle. This result, combined with the fact that
the remaining boundary conditions specify a fixed boundary zweibein,
imply that the action (\ref{Liou}) is indeed to be interpreted as a functional 
$I_{\rm reg}[\varphi]$ which generates correlation functions of the
boundary stress tensor.

We can now verify the result for the Weyl anomaly (\ref{weylA}), by
considering the variation of the generating functional $I_{\rm
reg}[\varphi]$ directly. We obtain, 
\be
  \left< T^i_i\right> = - 2 e^{-\varphi}\frac{\de I_{\rm
reg}[\varphi]}{\de \varphi} = - \frac{c}{24\pi} e^{-\varphi}
 \ptl\bar\ptl \varphi, 
\ee
which is equivalent to (\ref{weylA}) in the background
(\ref{zweibein}), where ${\cal R}^{(0)}=e^{-\varphi}\ptl\bar\ptl
\varphi$, and we have dropped the dependence on the
cosmological constant.

\subsubsection{Boundary light-cone gauge}

To exhibit the dependence on the chiral moduli $\mu$, we consider
the alternate boundary zweibein (\ref{lczwei})
\be
 e^{(0)} = \frac{1}{2}(dz + \mu d\bar z)J_- - \frac{1}{2}d\bar z J_+, 
 \label{lczb}
\ee 
which leads to the light-cone gauge metric (\ref{lcmet}). 
The decomposition of the group element for this case is conveniently
achieved by noting that (\ref{lczb}) may be obtained from
the zweibein (\ref{zweibein}) in conformal gauge by the 
quasi-conformal mapping,
\be
 \ph \rightarrow - \ln\ptl x, \;\;\;\;\;\;\; \bar\ph \rightarrow 0,
\ee
where $x$ satisfies the Beltrami-type equation (\ref{beltrami}),
while $\bar x = \bar z$,  and $\ptl$ now denotes $\ptl/\ptl z$. 

The potential term is unchanged under this mapping, so we concentrate
on the kinetic part. Evaluating (\ref{Ireg}) in this background,
we find,
\begin{equation}
 I_{reg}[x;y,\bar y] 
 = \frac{k}{8\pi}\int d^2 z \left[ 2\ln \ptl x (\ptl x \bar \ptl y - \bar
 \ptl x \ptl y + \ptl \bar y) + 4y \bar y \ptl x \right],
\label{Iregmu}
\end{equation}
and, again ignoring boundary terms, we can integrate 
out $y$ and $\bar y$. Making use of
(\ref{beltrami}) we find, 
$2y = -\ptl^2 x/(\ptl x)^2$ and $2\bar y = -\ptl \mu$,
leading to
\be
 I_{\rm reg}[\mu] = 
 -\frac{c}{48\pi} \int_{\Si} d^2 z 
 \frac{\ptl^2x}{\ptl x} \ptl\mu. \label{Ilc}
\ee
which we recognize as the light-cone gauge Polyakov action  $W_P[\mu]$
given in (\ref{chiral}).

One can generalize this approach to a more general background, but
these examples are sufficient for us to conclude that the dependence
on the boundary moduli is correctly encoded in the bulk action, 
consistent with the AdS/CFT correspondence. However, as mentioned
in Section~3, specifying the boundary metric does not uniquely
determine the bulk geometry due to the possibility for Brown-Henneaux
transformations. We will now explore the consequences of this 
for the boundary generating functional.

\subsection{Brown-Henneaux diffeomorphisms and stress 
tensor expectation values}

The consistency of the results of the preceding section hides 
the ambiguity of the Fefferman-Graham expansion regarding the
trace-free part of the first subleading term in the bulk metric. 
This ambiguity, equivalent to specifying the expectation value of the
boundary stress-tensor, is associated with the 
asymptotic symmetry of the space of bulk solutions under
Brown-Henneaux diffeomorphisms \cite{BH}, and the ambiguity can be
rephrased as specifying the position of a particular solution along
a Brown-Henneaux orbit.

The finite generalization of the mapping (\ref{inf}) corresponds
to 
\be
 T(z) \rightarrow (\ptl f)^2 T(f) - \frac{c}{24\pi} \{f,z\},\label{finite}
\ee
where $f(z)$ represents a local diffeomorphism of the circle in the
case that we start with a boundary of cylindrical topology, and
$\{f,z\}$ is the Schwarzian derivative,
\be
 \{f,z\} \equiv \frac{\ptl^3 f}{\ptl f} 
 - \frac{3}{2} \left(\frac{\ptl^2 f}{\ptl f}\right)^2. 
\ee
It is the Schwarzian term in the anomalous mapping (\ref{finite}) 
which leads to a finite charge \cite{BH} characterizing the 
position of a given solution on a Brown-Henneaux orbit. We will now
explore the consequences of this mapping for the results of the
previous section.

\subsubsection{Brown-Henneaux diffeomorphisms acting on the currents}

The Brown-Henneaux diffeomorphisms have a simple and direct action on
the Chern-Simons currents.  Let us go back to the expression for the currents 
(\ref{currents}) and impose the conditions $\alpha^3 = \bar \alpha^3$ 
(which are equivalent to the equations of motion for $y$ and $\bar
y$). In conformal coordinates $x=z$ and $\bar x = \bar z $, these 
conditions read $y=(1/2) \pp\varphi$ and $\bar y=(1/2) \bar\pp\varphi$
and the currents take the form,
\begin{equation}
\begin{array}{ll}
\alpha^+ =  {1 \over 2}e^{-\phi} \left[ T^\varphi dz + {\cal R} d\bar
  z\right]  \\
\alpha^- = e^{\phi} dz \\
\alpha^3 = - \pp\bar\phi \, dz + \bar\pp \phi \, d\bar z .  
\end{array}
\end{equation}
where ${\cal R}=\ptl\bar\ptl\varphi$ is the scalar coefficient of
the 2-form curvature of the background (\ref{zweibein}), and 
\be
 T^{\varphi} = \frac{24\pi}{c}\langle T \rangle = 
 \ptl^2\varphi - \frac{1}{2}(\ptl\varphi)^2 \label{Tliou}
\ee
can be identified with the Liouville stress-energy tensor 
of the conformal mode $\varphi$. Note that
similar formulae arise in the `anti-holomorphic' sector. 

Now consider the following local conformal transformations,
\begin{eqnarray}
z \rightarrow f(z),  & \ \ &  e^\phi \rightarrow e^\phi (\pp f)^{-1} 
  \nonumber\\
\bar z \rightarrow \bar f(\bar z),  &\ \  &  e^{\bar\phi} \rightarrow
e^{\bar\phi} (\bar\pp \bar f)^{-1}. \label{cmap} 
\end{eqnarray}
We observe that the conformal factor $\varphi$ then transforms 
appropriately as a 
Liouville field $ e^\varphi \rightarrow e^\varphi (\pp f)^{-1}(\bar\pp
\bar f)^{-1}$. 

Both $\alpha^-$ and $\alpha^3$ are trivially invariant under these 
transformations.  In order to analyze the transformation of $\alpha^+$
we note that the combination $e^{-\phi} dz$ has conformal dimension
$(-2,0)$ while $e^{-\phi} d\bar z$ has conformal dimension $(-1,-1)$.
Since $T^\varphi$ has conformal dimension $(2,0)$ and $\pp\bar\pp
\phi$ has conformal dimension $(1,1)$, the component $\alpha^+$
transforms anomalously via the Schwarzian derivative. This is
consistent with the discussion in Section~I on recognizing that
$\al^+$ enters the subleading term in the metric and encodes the 
boundary stress-energy tensor via $T^{\varphi}$. This can be made more
explicit by considering the form of the bulk metric.

\subsubsection{The general bulk metric and stress tensor expectation values}  

We can insert the expressions for the currents into (\ref{fullmet}) to
obtain the corresponding form of the bulk metric. We find,
\bea
 ds^2_{3D} &=& dr^2 - e^{2r+\varphi}dxd\bar x
  + \frac{1}{2} T^{\varphi} dx^2 +  \frac{1}{2} \bar T^{\varphi} d\bar x^2
  + {\cal R}dxd\bar x \nonumber \\
  && \;\;\;\;\;\;\;\;\; - \frac{1}{2}e^{-2r-\varphi}(T^{\varphi}dx +
 {\cal R}d\bar x)(\bar T^{\varphi}d\bar x +
 {\cal R}dx), \label{fmet}
\eea
where, from the subleading terms, we identify the general form of the
expansion given in (\ref{FGgen}).
Indeed, the metric (\ref{fmet}) is the general solution associated with
a conformally flat boundary geometry, as determined recently in
\cite{RS2,K2}. However, the mapping (\ref{cmap}) implies that 
the metric (\ref{fmet}), in which there is a direct 
correspondence between the subleading terms and the Liouville 
stress-energy tensor for the representative boundary metric,
is not completely general. The fact that $T^{\varphi}$ transforms
anomalously is conveniently encoded in (\ref{fmet}) by writing it 
in a new coordinate system
$\{x,\bar x\}\rightarrow \{z=f(x),\bar z=\bar{f}(\bar x)\}$, 
\bea
 ds^2_{3D} &=& dr^2 - e^{2r+\varphi}dzd\bar z
  + \frac{1}{2} (T^{\varphi}+\tilde{T})dz^2 +  \frac{1}{2} (\bar
T^{\varphi}+\tilde{\bar T}) d\bar z^2
  + {\cal R}dzd\bar z \nonumber \\
  && \;\;\;\;\;\;\;\;\; 
 - \frac{1}{2}e^{-2r-\varphi}((T^{\varphi}+\tilde{T})dz +
 {\cal R}d\bar z)((\bar T^{\varphi}+\tilde{\bar T})d\bar z +
 {\cal R}dz), \label{fmet2}
\eea
where $\ptl = \ptl /\ptl z$, and $\tilde{T}=\tilde{T}(z)$ is an arbitrary
chiral function given by $\tilde{T}(z)=-c\{x,z\}/24\pi$. 
This representation makes explicit the anomalous shift (\ref{finite})
of the subleading terms under Brown-Henneaux diffeomorphisms, 
in accordance with the mapping $T(z) = \langle T_{\rm CFT} \rangle$
to the dual CFT \cite{martinec,banados99,adsrev,bers,eff}.

In order to understand how the generating functional is consistent
with this shift (via (\ref{Treln})), it is convenient to consider
a particular example. Specifically, setting $\varphi$ 
to zero in (\ref{fmet2}), the metric reduces to (dropping the tildes)
\bea
 ds^2_{3D} &=& dr^2 - \left(e^{2r} + e^{-2r}T\bar T\right)dzd\bar z
  + \frac{1}{2} Tdz^2 +  \frac{1}{2} \bar Td\bar z^2, \label{fmet3}
\eea
which is equivalent to the general solution with a flat boundary 
geometry obtained in \cite{banados99}. In interpreting (\ref{Treln})
for such a background it is important to recall that the boundary metric
is only specified up to a Weyl rescaling. Thus, the actual boundary
metric in (\ref{fmet3}) is any geometry in the same conformal class as
the flat one. From (\ref{fmet2}), we see that the ambiguity 
reflects additive `free field' shifts in the conformal mode $\varphi$. This 
explains why the generating functional is actually
nonvanishing for (\ref{fmet3}) and leads on variation, via (\ref{Treln}), to
nonzero expectation values for $\langle T_{\rm CFT}\rangle$ and 
$\langle {\bar T}_{\rm CFT}\rangle$. Following \cite{martinec}, 
this interpretation of the allowed free-field shifts in $\varphi$
may also be understood in terms of the connection between the oscillator modes
in an expansion of the Liouville field $\varphi$, and the bulk
`gravitational wave' perturbations specified by $T$ and $\bar T$.

In the discussion above, we concentrated for simplicity on the
boundary conformal gauge. If we instead consider boundary 
geometries in the conformal class of the light-cone metric
(\ref{lcmet}), then the situation is similar with the complication
that local conformal transformations on the boundary amount to
`chiral' free-field shifts in the conformal mode. The stress-energy 
tensor, $T(x)= c\{x,z\}/24\pi$ given by the variation
of (\ref{Ilc}), then shifts according to (\ref{finite}) via the Schwarzian
which is now chiral with respect to the light-cone structure
determined by $\mu$ (see e.g. \cite{verlinde}).

\section{Global issues and concluding remarks}
  
Making use of the Chern-Simons formalism, we have shown that the
dependence of the regularized bulk gravitational action on local moduli of
the boundary metric is precisely as one would expect for the Polyakov
action, consistent with the AdS/CFT correspondence.
In this final section we comment on some of the important features
that were ignored in the analysis of Section~IV, specifically global
data associated with the holonomy of the gauge fields.

One of the main motivations for this work was to understand the
manifestation of the data specifying the bulk solutions
in the boundary effective action. In 3D, we can roughly characterize 
this bulk data in terms of Brown-Henneaux `gravitational waves' 
at the boundary, and global data conveniently described by 
holonomies of the Chern-Simons gauge field. The latter can
alternatively be thought of as the modes associated with `large'
Brown-Henneaux diffeomorphisms which change the topology.
The construction of the boundary effective action in Section~IV,
in contrast to earlier analyses \cite{CHvD,RS2,K2} which made use of the
WZW model at intermediate stages\footnote{As noted earlier, the work
of Coussaert et al. \cite{CHvD} proceeded along different lines. 
Nonetheless, we can consider
using our formulation to obtain the analogue of the CHvD action \cite{CHvD}
by solving the constraints, with appropriate fall-off conditions for
the fields at infinity. In doing this, an important 
distinction is that we need to identify their Liouville field 
$\phi$ with the B\"acklund transform of the field $\varphi$ appearing 
here (provided we choose $\la^2=0$ as the renormalization condition).
i.e. $\ptl \varphi = -\ptl \ph + e^{(\ph-\varphi)/2}$, $\bar \ptl \varphi =
\bar\ptl \ph - e^{(\ph+\varphi)/2}$ \cite{D'HJ}. This is due to the
use of the Polyakov-Wiegmann identity in \cite{CHvD} which performs
the appropriate canonical transformation, and this mapping then explains
the form of the bulk metric which can be determined in 
the approach of \cite{CHvD}.}, should allow a more direct analysis
of the dependence of the effective action on the global bulk 
data (see also \cite{wzwredux,globalredux}). We plan
to report on this elsewhere. However, some comments are in order
regarding the need for this extension of the effective action 
in the analysis of Section~IV.

Recall that, in general, the action (\ref{Ifinite})
can only be localized once a suitable choice of the holonomies 
tr$(\exp\oint A)$ and tr$(\exp\oint \bar A)$ is made, thus
specifying the topology. Within SL(2,$\Re$), these fall into
three conjugacy classes: elliptic holonomies, conjugate to rotations, 
correspond to conical singularities; hyperbolic holonomies, 
conjugate to dilations, correspond to non-extreme back holes; 
while extreme black holes are associated with parabolic 
holonomies conjugate to translations \cite{BHTZ}. Including this global data
will be important in extending the considerations both of this paper,
and of \cite{CHvD}, to bulk and boundary geometries of more
complicated topology. 

As emphasized in \cite{martinec}, the 
classification of bulk conjugacy classes is mirrored in the classical 
solutions of Liouville theory, or in other words the uniformization of
the boundary $\Si$. Thus one suspects that the
holonomies map to zero modes of the Liouville field, and it would
be interesting to understand how this applies in the context of the
generating functional studied here, where the Liouville field in
question can be associated with the bulk radial scale rather than a
dynamical field as in \cite{CHvD}. In practice, this picture is 
complicated because in generic cases there is no global Liouville
field corresponding to the bulk solution \cite{Navarro-N,HMS,RS1}.
A straightforward example is given by a spinning black hole. For
nonzero angular momentum, the corresponding Liouville solution can
only be specified locally. It is interesting to note
in this regard that the existence of global solutions for the 
Liouville field is closely connected with the existence of
Killing spinors \cite{RS1}. The absence of a global
Liouville solution for the spinning black hole then correlates with
the absence of Killing spinors for the 
generic BTZ solution \cite{HvD}. This suggests that the 
holonomies of bulk solutions will be an important ingredient in 
mapping out the zero-mode structure of the Liouville field,
and consequently the global structure of the geometry seen by the
dual CFT.

\bigskip

\centerline{\bf Acknowledgements}
MB thanks M. Henneaux for a useful conversation, and   
AR thanks the Departamento de F\'{\i}sica, P. Universidad Cat\'olica 
de Chile, where part of this work was completed, for warm hospitality.
This work was supported in part by FONDECYT (Chile) Grants \# 1000744,
7000744, and 3000026.


\end{document}